\documentclass[11pt]{article}
\usepackage{amsmath,amssymb}

\def\a{\alpha}
\def\da{\dot\alpha}
\def\b{\beta}
\def\db{\dot\beta}

\def\d{\delta}

\def\e{\epsilon}           
\def\f{\phi}               

\def\m{\mu}
\def\n{\nu}
  
\def\q{\theta}    \def\bq{\bar\theta}                
\def\s{\sigma}                                   

\def\pa{\partial}              

\begin{document}
\thispagestyle{empty}
\null\vskip-24pt \hfill CERN-TH/2003-180 \vskip-10pt \hfill LAPTH-992/03

\begin{center}
\vskip 0.2truecm {\Large\bf
Non-anticommutative $N=2$ super-Yang-Mills theory with singlet deformation \vskip
0.0truecm }

\vskip2cm {\Large  Sergio Ferrara$^\dagger{}^\sharp$ and Emery 
Sokatchev$^\ddagger$ }\\ \vskip6mm {\it e-mail: Sergio.Ferrara, Emery.Sokatchev@cern.ch} \\ \vspace{1cm} $^\dagger$ CERN 
Theoretical Division, CH 1211 Geneva 23, Switzerland 
\\ \vspace{6pt}
$^\sharp$ Laboratori Nazionali di Frascati, INFN, Italy                 
\\ \vspace{6pt}

$^\ddagger$ Laboratoire d'Annecy-le-Vieux de Physique 
Th\'{e}orique\footnote[1]{UMR 5108 associ{\'e}e {\`a} 
 l'Universit{\'e} de Savoie} LAPTH, Chemin
de Bellevue - BP 110 - F-74941 Annecy-le-Vieux Cedex, France       
\end{center}

\vskip 1truecm \Large
\centerline{\bf Abstract} \normalsize We consider a non-anticommutative $N=2$ superspace with an SU(2) singlet and Lorentz scalar deformation parameter, $\{\q^{\a i},\q^{\b j}\}_\star = -2iP \e^{\a\b}\e^{ij}$. We exploit this unique feature of the $N=2$ case to construct a deformation of the non-Abelian super-Yang-Mills theory which {\it preserves the full $N=2$ supersymmetry} together with the SU(2) R symmetry and Lorentz invariance. The resulting action describes a kind of ``heterotic special geometry" with antiholomorphic prepotential $\bar f(\bar\f) = {\rm Tr}\left( \bar\f^2 (1+P\bar\f)^{-2} \right)$.

\newpage
\setcounter{page}{1}\setcounter{footnote}{0}

\section{Introduction}

Supersymmetric field theories in non-commutative superspace have recently attracted considerable interest in view of the fact that they describe some superstring effective actions in the background of space-time supergravity fields \cite{CDS}-\cite{SvN}. Both bosonic \cite{SW} and fermionic \cite{FL,KPT,S} deformations of superspace have been considered, giving rise to non-commutative ($[x,x]\neq0$) and non-anticommutative ($\{\q,\q\}\neq0$) coordinates. In the latter case, depending on whether one chooses the supercovariant derivatives $D_\a$ or the supersymmetry generators $Q_\a$ as the differential operator defining the Poisson bracket, one obtains a supersymmetric \cite{FL,KPT,FLM} or a partially supersymmetric \cite{S} Moyal-Weyl star product.

In the case of extended $(N>1)$ supersymmetry there exists a wider class of non-anticommutative deformations of superspace \cite{FLM}:
\begin{equation}\label{1}
  \{\q^{\a i},\q^{\b j}\}_\star = P^{(\a\b)(ij)} +  \e^{\a\b}P^{[ij]}\,.
\end{equation}
Here $\a=1,2$ is a left-handed Lorentz spinor index; $i=1,\ldots,N$ is an index of the fundamental irrep of the R symmetry group  SU($N$); $\star$ denotes the Moyal-Weyl star product; $P^{(\a\b)(ij)} = \s_{\m\n}^{\a\b}\; P^{\m\n\; (ij)}$ is a constant self-dual Lorentz tensor symmetric in the SU($N$) indices, while  $P^{[ij]}$ is a scalar antisymmetric in the SU($N$) indices. The case $N=2$ is exceptional since the second deformation parameter in (\ref{1}) becomes an SU(2) singlet, $P^{[ij]} = -2iP\e^{ij}$. Setting $P^{(\a\b)(ij)}=0$ and keeping only $P$ is the unique way of deforming superspace without breaking the R symmetry (more precisely, its SU(2) part) and Lorentz invariance.

In this paper we investigate the deformed $N=2$ super-Yang-Mills theory in a non-anticommutative superspace with a singlet scalar deformation parameter $P$. We show that it is possible to define the star product through the spinor covariant derivatives $D_{\a i}$ with the striking result that the deformed super-Yang-Mills theory preserves not only the SU(2) and Lorentz symmetries, but also the full $N=2$ supersymmetry (or $(1,1)$ supersymmetry in  Euclidean notation). This should be compared to the recently proposed deformations of the $N=1$ and $N=2$ gauge theories which preserve only the $(1/2,0)$ half of the $(1/2,1/2)$ supersymmetry \cite{S} and the $(1/2,0)$ quarter of the $(1,1)$ supersymmetry \cite{BSe,AIO}, respectively.

\section{Star products in $N=2$ superspace: Chirality vs G-analyticity}

Four-dimensional $N=2$ superspace is parametrized by an even (commuting) four-vector $x^\mu$ and by an SU(2) doublet of odd (anticommuting) Lorentz spinor coordinates $\q^{\a i},\bq_{\da i}$, $i=1,2$. As discussed in the Introduction, we wish to study the following non-anticommutative deformation:
\begin{equation}\label{2}
  \{\q^{\a i},\q^{\b j}\}_\star = -2iP \e^{\a\b}\e^{ij}\,
\end{equation}
(the coefficient $-2i$ is chosen for convenience). It should be stressed that this definition violates reality, i.e., $\bq \neq (\q)^*$ anymore. This is only possible in Euclidean space, but following \cite{S}, we will continue to use the Lorentzian signature notation.

In different terms, the non-anticommutativity of the $\q$s is a consequence of replacing the usual product $A(\q)B(\q)$ by the fermionic Moyal-Weyl star product \cite{FL,KPT}
\begin{equation}\label{3}
  A\star B = A  \exp\left(iP \e^{\a\b}\e^{ij} \overleftarrow{\eth}_{\a i} \overrightarrow{\eth}_{\b j} \right) B
\end{equation}
Here $\eth_{\a i}$ is a differential operator such that $\{\eth_{\a i},\eth_{\b j}\} = 0$ and $\eth_{\a i} \q^{\b j} = \pm\d^\b_\a \d^j_j$. The anticommutativity of the $\eth_{\a i}$ guarantees that the star product is associative, 
\begin{equation}\label{4}
  (A\star B)\star C = A\star (B\star C)\,,
\end{equation}
but it remains, in general, non-commutative.   

As discussed in detail in Ref. \cite{FLM}, there exist two inequivalent choices of $\eth_{\a i}$: One can use either the supersymmetry generators $Q_{\a i}$ or the covariant spinor derivatives $D_{\a i}$. The first choice has the advantage that it preserves chirality, a basic property of $N=1$ supersymmetric field theory, but it breaks half of the supersymmetry (since $\{\bar Q, Q\} \neq 0$). This choice was successfully used in Ref. \cite{S} to construct deformations of the $N=1$ theories of matter and gauge supermultiplets which preserve the left-handed (i.e., generated by $Q_\a$) half of $N=1$ supersymmetry. The second choice preserves the full supersymmetry (since $\{\bar Q, D\} = \{Q,D\} = 0$) but it breaks left-handed chirality (defined by $\bar D \Phi=0$).

The difference between these two choices becomes more clear after examining the deformed (anti)commutators of the coordinates in the chiral basis in superspace. Chiral superfields are defined by the differential constraint
\begin{equation}\label{8}
  \bar D^i_{\da}\Phi(x,\q,\bq) = 0
\end{equation}    
where the supercovariant derivative $\bar D^i_{\da}$ and its conjugate $D_{\a i}$ satisfy the algebra\footnote{We use the conventions of \cite{book}. In particular, we raise and lower SU(2) and Lorentz spinor indices with the help of the $\e$ tensor, e.g., $D^i = \e^{ij}D_j$, $\bar D_i = \e_{ij}\bar D^j$ ($\e_{ij}\e^{jk} = \d^k_i$, $\e_{12}=1$).}
\begin{eqnarray}
  &&\{D_{\a i},D_{\b j}\} = \{\bar D^i_{\dot\alpha},\bar D^j_{\dot\beta}\} =
 0\,, \nonumber\\
  &&\{D_{\a i},\bar D_{\dot\alpha}^j\} = 2i\delta^j_i
 (\sigma^\m)_{\alpha\dot\alpha}\pa_\m\,. \label{ADalgebra}
\end{eqnarray}
Chirality becomes manifest in the appropriate left-handed chiral basis in superspace
\begin{equation}\label{9}
  x_L^\m = x^\m + i \q^i\s^\m\bq_i, \quad \q^{\a i}, \quad \bq^{\da}_i\,,
\end{equation}
in which $\bar D^i_{\da} = \pa/\pa\bq^{\da}_i$ and condition (\ref{8}) has the solution $\Phi = \Phi(x_L,\q)$. In this new basis the left-handed supersymmetry generator $Q_{\a i}$ and supercovariant derivative $D_{\a i}$ take the form
\begin{equation}\label{chQD}
  Q_{\a i} = -i\frac{\pa}{\pa \q^{\a i}}\,, \qquad D_{\a i} = -\frac{\pa}{\pa \q^{\a i}} +2i \bq^{\da}_i (\s^\m)_{\a\da}\frac{\pa}{\pa x^\m_L}\,.
\end{equation}
Then, using the star product (\ref{3}) with $\eth_{\a i} = iQ_{\a i}$ or alternatively with $\eth_{\a i} = D_{\a i}$ to compute the (anti)commutators of the coordinates (\ref{9}), we find that the $Q$ deformation only affects the $\q$s as postulated in (\ref{2}),
\begin{equation}\label{Qch}
   \{\q^{\a i},\q^{\b j}\}_{\star Q} = -2iP \e^{\a\b}\e^{ij}\,, 
\end{equation}
while the $D$ star product deforms the entire left-handed chiral subspace $x^\m_L,\q^{\a i}$:
\begin{eqnarray}
 [x^\m_L,x^\n_L]_{\star D} &=& 8P \bq_i\tilde\s^{\m\n} \bq^i \nonumber\\
 {[}x^\m_L,\q^{\a i}]_{\star D} &=& -4P (\bq^i\tilde\s^{\m})^\a \label{Dch}\\
 \{\q^{\a i},\q^{\b j}\}_{\star D} &=& -2iP \e^{\a\b}\e^{ij}\,. \nonumber
\end{eqnarray}

In $N=2$ supersymmetry we have an alternative choice of an invariant subspace involving half of the odd coordinates, the so-called Grassmann (or G-) analytic superspace \cite{GIO}. It is best formulated by introducing new  even superspace coordinates, the so-called harmonic variables $u^\pm_i$ \cite{GIKOS,book} which form an SU(2) matrix:
\begin{equation}
\parallel u \parallel\ \in {SU}(2):\qquad u^{+i}u^-_i \equiv u^{+i}\e_{ij}u^{-j} = 1\,,\qquad
\overline{u^{+i}} = u^-_i\,. \label{5}
\end{equation} 
With the help of $u^\pm_i$ we can split the SU(2) doublet indices of the spinor derivatives into U(1) projections,
\begin{equation}\label{A3.104}
 D^\pm_{\alpha} = u^\pm_i D^i_{\alpha}, \qquad \bar D^\pm_{\dot\alpha} =
 u^\pm_i \bar D^i_{\dot\alpha}\,,
\end{equation}
so that, e.g., $D_{\a i} = u^+_i D_{\a}^- - u^-_i D_{\a}^+$ (using the completeness condition $u^+_iu^-_j - u^+_ju^-_i = \e_{ij}$). 
In this way we see that the algebra (\ref{ADalgebra}) contains the ideal
\begin{equation}\label{6}
  \{D^+_\alpha, D^+_\beta\} = \{D^+_\alpha, \bar D^+_{\dot\alpha}\} =
\{\bar D^+_{\dot\alpha}, \bar D^+_{\dot\beta}\} = 0\,,
\end{equation}
which allows us to define {\it G-analytic superfields} satisfying the conditions
\begin{equation}\label{7}
  D^+_\a\Phi(x,\q,\bq,u) = \bar D^+_{\da}\Phi(x,\q,\bq,u) = 0\,.
\end{equation}
This is the natural generalization of the notion of chirality
in extended ($N>1$) supersymmetry. In fact, in the case $N=2$ G-analyticity takes over the fundamental r\^ole which chirality plays in $N=1$ supersymmetric field theory (see Section 3 for a brief review). 

Just as chirality becomes manifest in the chiral basis (\ref{9}), G-analyticity (\ref{7})  does so in the appropriate G-analytic basis
\begin{equation}\label{10}
  x_A^\m = x^\m - i \q^i\s^\m\bq^j (u^+_iu^-_j + u^+_ju^-_i), \quad \q^{\pm\a} = \q^{\a i}u^\pm_i, \quad \bq^{\pm\da} = \bq^{\da i}u^\pm_i\,, 
\end{equation}
where
\begin{equation}\label{10'}
  D^+_\a = \pa/\pa\q^{-\a}\,, \qquad \bar D^+_{\da} = \pa/\pa\bq^{-\da}
\end{equation}
and conditions (\ref{7}) are solved by $\Phi = \Phi(x_A,\q^+, \bq^+,u)$. In other words, $N=2$ superspace has a G-analytic subspace closed under supersymmetry which involves only half (but not the chiral half!) of the odd coordinates, $\q^+, \bq^+$.

Let now us see how the two deformations, $Q$ and $D$, affect the coordinates in the G-analytic basis (\ref{10}). In this basis we have
\begin{eqnarray}
  Q^+_\a &=& Q^i_\a u^+_i = i\frac{\pa}{\pa \q^{-\a}} +2 \bq^{+\da} (\s^\m)_{\a\da}\frac{\pa}{\pa x^\m_A} \nonumber\\
  Q^-_\a &=& Q^i_\a u^-_i = -i\frac{\pa}{\pa \q^{+\a}} \label{QDan}\\
  D^-_\a &=& D^i_\a u^-_i = -\frac{\pa}{\pa \q^{+\a}} +2i \bq^{-\da} (\s^\m)_{\a\da}\frac{\pa}{\pa x^\m_A} \nonumber
\end{eqnarray}
Next, projecting the operator $\eth_{\a i}$ in the star product with harmonics and using (\ref{QDan}), we find the effect of the $Q$ and $D$ deformations:
\begin{eqnarray}
  &&{[}x^\m_A,\q^{+\a}]_{\star Q} = -4P (\bq^+\tilde\s^\m)^\a\,, \qquad \{\q^{+\a},\q^{-\b}\}_{\star Q} = 2iP\e^{\a\b}\,; \label{defQan}\\
  &&{[}x^\m_A,\q^{-\a}]_{\star D} = -4P (\bq^-\tilde\s^\m)^\a\,, \qquad \{\q^{+\a},\q^{-\b}\}_{\star D} = 2iP\e^{\a\b}\,. \label{defDan}
\end{eqnarray}
Besides the desired non-anticommutativity of the odd coordinates (the non-vanishing harmonic projection of eq. (\ref{2})), the two deformations also induce a non-trivial commutator among the even and odd coordinates. However, with $Q$ this new deformation takes place within the G-analytic subspace itself, while with $D$  it stays outside ($\q^-,\bq^-$ do not belong to  the G-analytic subspace).

The discussion above shows that before using the singlet scalar deformation parameter $P$ in $N=2$ superspace, we have first to decide whether we wish to preserve the notion of chirality or of G-analyticity. In the former case it is preferable to work with the $Q$ deformation, while in the latter the $D$ deformation should be our choice. Since both the $N=2$ matter and gauge theories are based on G-analyticity, from now on we adopt the definition 
\begin{equation}\label{26}
  A\star B = A  \exp\left(iP \e^{\a\b}\e^{ij} \overleftarrow{D}_{\a i} \overrightarrow{D}_{\b j} \right) B\,.
\end{equation}

\section{$N=2$ matter and gauge theories} 

In this section we give a very brief review of the harmonic superspace formulation of the $N=2$ matter and gauge theories (for the details see \cite{book}). In both cases the basic objects, the $N=2$ hypermultiplet and the gauge potential, are described by G-analytic superfields. It is important to make it clear that because of the harmonic-dependent shift (\ref{10}) a G-analytic superfield $\Phi(x_A,\q^+, \bq^+,u)$ is necessarily a function of the harmonic variables. Such functions are defined as harmonic expansions on the coset $S^2 \sim $ SU(2)/U(1). In practical terms, this expansion goes over the irreducible monomials of $u^\pm$ carrying a definite U(1) charge $q$. For example, for $q\geq 0$ we define
\begin{equation}
f^{(q)}(u) = \sum^\infty_{n=0} f^{(i_1\ldots i_{n+q}j_1\ldots
j_n)} u^+_{i_1}\ldots u^+_{i_{n+q}}u^-_{j_1}\ldots u^-_{j_n}\; ,
\label{11}
\end{equation}
where the coefficients $f^{(i_1\cdots j_n)}$ are symmetric traceless SU(2) tensors (irreps of isospin $n/2$). These infinite expansions can be reduced to polynomials by the differential condition of harmonic (or H-) analyticity
\begin{eqnarray}
  \pa^{++}f^{q}(u) = 0&\Rightarrow& f^{q}(u) = f^{i_1\cdots i_q}u^+_{i_1}\cdots u^+_{i_q} \ \mbox{if $q\geq0$;} \label{12}\\
  &\Rightarrow& f^{q}(u) = 0  \ \mbox{if $q<0$,}\label{12'}
\end{eqnarray}
where the harmonic derivative $\pa^{++}$ acts as follows:
\begin{equation}\label{13}
   \pa^{++} u^- = u^+\,, \quad \pa^{++}u^+ = 0\,.
\end{equation}
This derivative and its conjugate defined by
\begin{equation}\label{13'}
   \pa^{--} u^- = 0\,, \quad \pa^{--}u^+ = u^-\,,
\end{equation} 
are just the raising and lowering operators of the SU(2) algebra realized on the charges $\pm$ of the harmonics:
\begin{equation}\label{14}
  [\pa^{++},\pa^{--}] = \pa^0\,.
\end{equation}
Here $\pa^0$ is the charge counting operator, $\pa^0 u^\pm = \pm u^\pm$ and $\pa^0 f^{q}(u) = q f^{q}(u)$.

In the G-analytic basis (\ref{10}), where the spinor derivatives $D^+_\a, \bar D^+_{\da}$ become short (see (\ref{10'})), the harmonic derivatives acquire new, space-time terms:
\begin{equation}\label{15}
  D^{\pm\pm} = \pa^{\pm\pm} - 2i \q^\pm\s^\m \bq^\pm \frac{\pa}{\pa x^\m_A}\,.
\end{equation}
As a consequence, the combination of G- and H-analyticity on a harmonic superfield of charge $+1$ is equivalent to the on-shell constraints of the $N=2$ hypermultiplet \cite{GIKOS,book}:
\pagebreak
\begin{equation}\label{16}
  D^{++} q^+(x_A,\q^+, \bq^+,u) = 0 \ \Rightarrow
\end{equation}
$$
q^+ = f^i(x_A)u^+_i +\theta^{+\alpha}\psi_\alpha(x_A)
+\bar\theta^+_{\dot\alpha}\bar\kappa^{\dot\alpha}(x_A) +2i\theta^+\sigma^\m\bar\theta^+ \partial_\m f^i(x_A)u^-_i\,,
$$
where the component fields satisfy their free field equations $ \square f^i=\pa\cdot\s\psi
=\pa\cdot\s\bar\kappa=0$. Most importantly, this approach allows one to go off shell and write down the free hypermultiplet action in the form of a G-analytic superspace integral:
\begin{equation}\label{17}
  S^{N=2}_{{\rm matter}} = -\int du\, d^4x_A\, d^2\q^+\, d^2\bq^+ \; \tilde q^{+}D^{++}q^+ \,.
\end{equation}
Note that the conjugate superfield $\tilde q^{+}(x_A,\q^+, \bq^+,u)$ is also G-analytic and not antianalytic.\footnote{In $N=2$ harmonic superspace there exists an operation combining complex conjugation with the antipodal map on $S^2$ which preserves G-analyticity \cite{book}.} This is an important difference from the $N=1$ case where the free matter action is written down as a full superspace integral of a chiral and an antichiral superfields, $S^{N=1}_{{\rm matter}} = \int  d^4x\, d^2\q\, d^2\bq \; \bar\Phi(x_R,\bq)\Phi(x_L,\q)$ (here $x_R$ is the complex conjugate of ${x_L}$ from (\ref{9})).

The form of the hypermultiplet action (\ref{17}) immediately suggests to introduce the $N=2$ gauge superfield as the {\it gauge connection for the harmonic derivative}. Suppose that the matter superfield $q^+$ transforms in, e.g., the adjoint representation of a non-Abelian gauge group, $\d q^+ = i[\Lambda, q^+]$. It is clear that the gauge parameter must be G-analytic, $\Lambda = \Lambda(x_A,\q^+, \bq^+,u)$, just like the matter superfield itself.\footnote{Similarly, in the $N=1$ case the (anti)chiral matter superfields transform with  (anti)chiral gauge parameters.} Then the harmonic derivative in (\ref{17}) needs to be covariantized:
\begin{equation}\label{18}
  D^{++} q^+ \ \rightarrow \ {\nabla}^{++} q^+ = D^{++} q^+ + i [V^{++}, q^+]\,,
\end{equation}
where the gauge connection is another G-analytic superfield  $V^{++}(x_A,\q^+, \bq^+,u)$. It transforms in the familiar way
\begin{equation}\label{19}
  \d V^{++} = -D^{++}\Lambda + i[\Lambda, V^{++}]\,.
\end{equation}
This harmonic connection is the unconstrained off-shell gauge superfield of $N=2$ SYM theory. Notice the sharp difference from the $N=1$ case, where the gauge superfield has the form of a gauge transformation $e^{iV(x,\q,\bq)}$ rather than of a gauge connection.

The component content of the SYM theory is revealed in the Wess-Zumino gauge. It is obtained by examining the linearized (or Abelian) gauge transformation $\d V^{++} = -D^{++}\Lambda$ and by comparing the component content of the gauge superfield and parameter. Unlike the $N=1$ case, where the gauge superfield $V(x,\q,\bq)$ is real and the parameters are (anti)chiral, here both of them are of the same, G-analytic type. However, the difference comes from their U(1) charges, $+2$ for $V^{++}$ and $0$ for $\Lambda$. For instance, the lowest components in their $\q$ expansions are (recall (\ref{11})) $V^{++}|_{\q=0} = v^{ij}(x_A)u^+_iu^+_j + \cdots$ and $\Lambda|_{\q=0} = a(x_A) + a^{ij}(x_A)u^+_iu^-_j + \cdots$. Thus, we can completely gauge away $V^{++}|_{\q=0}$ by using the entire $\Lambda|_{\q=0}$ but its SU(2) singlet component $a(x_A)$ which is identified with the ordinary gauge parameter. Proceeding in the same way with all the components in the two Grassmann and harmonic expansions, it is not hard to show \cite{book} that the following WZ gauge exists:
\begin{eqnarray}\label{20}
 V^{++}_{\rm WZ}  &=& -2i\theta^+\sigma^\m\bar\theta^+
A_\m(x_A) - i \sqrt2 (\theta^+)^2\bar \phi(x_A) + i\sqrt2(\bar\theta^+)^2 \phi(x_A)
\nonumber \\ && +\,4(\bar\theta^+)^2\theta^{+\alpha}\lambda^i_{\alpha}(x_A) u^{-}_i
-4(\theta^+)^2\bar\theta^+_{\dot\alpha}\bar\lambda^{\dot\alpha i }(x_A) u^{-}_i + 3(\theta^+)^2(\bar\theta^+)^2 D^{ij}(x_A)u^-_iu^-_j\,; \nonumber \\
\Lambda_{\rm WZ} &=& a(x_A)\,.
\end{eqnarray}
Clearly, in the WZ gauge the connection $V^{++}$ contains only a finite set of components, the gauge field $A_\m$, the physical scalars $\f, \bar\f$ and spinors $\lambda,\bar\lambda$ and the auxiliary field $D^{ij}$. The rest of the infinite harmonic expansion on $S^2$ are pure gauge degrees of freedom which are matched by the expansion of the parameter $\Lambda$. Note also that $V^{++}_{\rm WZ}$ is nilpotent, just like the $N=1$ $V_{\rm WZ}$.

In the standard Minkowski superspace formulation of the $N=2$ gauge theory the G-analytic superfield $V^{++}$ satisfies a reality condition, $\widetilde{V^{++}} = V^{++}$. Consequently, the components in (\ref{20}) are either real ($A_\m$ and $D^{ij}$) or are related to each other by complex conjugation ($\f$ to $\bar\f$ and $\lambda$ to $\bar\lambda$). However, in view of the deformation (\ref{26}) which we want to introduce, we should use a Euclidean formulation where $\bq$ is not the complex conjugate of $\q$. In this case the scalars $\f$, $\bar\f$ and the spinors $\lambda$, $\bar\lambda$ should be treated as {\it independent fields} and not as complex conjugates. This will be important in our discussion of the deformed SYM action in Section 4.

The $N=2$ SYM action can be written down as an {\it (anti)chiral superspace} integral:
\begin{equation}\label{21}
  S^{N=2}_{\rm SYM} =\frac{1}{64}\int d^4x_L\, d^4\q\; \mbox{Tr} ( W W) =\frac{1}{64}\int d^4x_R\, d^4\bq\; \mbox{Tr} (\overline{W} \overline{W})\,.
\end{equation}
The curvature superfield $\overline{W}$ arises in the anticommutator of two gauge covariant spinor derivatives, $\{ {\nabla}_{\alpha}^i, {\nabla}_{\beta}^j\} = -
\frac{i}{2}\epsilon^{ij} \epsilon_{\alpha\beta} \overline{W}$. In fact, this is the only (anti)chiral object allowed in the $N=2$ non-Abelian gauge theory (the chirality condition on any other superfield would be incompatible with the algebra of the gauge covariant derivatives). In the harmonic superspace approach $\overline{W}$ is not expressed directly in terms of the unconstrained gauge superfield $V^{++}$, but in terms of the gauge connection for the other harmonic derivative $D^{--}$ (recall (\ref{15})), ${\nabla}^{--} = D^{--} + i V^{--}$. Although the pure harmonic derivatives (\ref{13}), (\ref{13'}) are related by complex conjugation, this is not true for their supersymmetrized and gauge covariant counterparts. Instead, the condition which relates $V^{--}$ to the G-analytic prepotential $V^{++}$ is the covariant version of the SU(2) commutation relation (\ref{14}) (the charge operator $\pa^0$ remains flat):
\begin{equation}\label{22'}
  [{\nabla}^{++},{\nabla}^{--}] = \pa^0 \ \Rightarrow \ D^{++}V^{--} - D^{--}V^{++} + i\left[ V^{++}, V^{--}\right] = 0\,,
\end{equation}
which can also be rewritten as follows:
\begin{equation}\label{22}
  {\nabla}^{++} V^{--} = D^{--}V^{++}\,.
\end{equation}
Note that  $V^{--}$ cannot be G-analytic since the derivative $D^{--}$ does not commute with $D^+, \bar D^+$.
This is a differential equation for $V^{--}$ on $S^2$ which has a unique solution.\footnote{The reason is that the homogeneous equation $D^{++}V^{--} = 0$ for a negative-charged harmonic function has no solution, see (\ref{12'}).} The explicit solution, which will not be needed here, can be represented as a power series in $V^{++}$ \cite{Z,book} or it can be worked out directly in the WZ gauge (\ref{20}).

Once $V^{--}$ has been found, the (anti)chiral curvatures $W, \overline{W}$ are given by the simple expressions
\begin{equation}\label{23}
  W =  \bar D^+_{\da} \bar D^{+\da} V^{--} \equiv (\bar D^+)^2  V^{--}\,, \qquad \overline{W} =  D^{+\a} D^+_\a V^{--}  \equiv (D^+)^2  V^{--} \,.
\end{equation}
Let us check the gauge covariance of, e.g., $\overline{W}$:
\begin{eqnarray}
  \d \overline{W} &=&  (D^+)^2 \left(-D^{--}\Lambda + i[\Lambda, V^{--}]\right)\nonumber\\
  &=& D^{+\a} D^-_\a \Lambda + i[\Lambda, (D^+)^2V^{--}]\nonumber\\
  &=& i[\Lambda, \overline{W}]\,.  \label{24}
\end{eqnarray}
Here  we have used the G-analyticity of the gauge parameter $\Lambda$ and the flat (anti)commutation relations
\begin{eqnarray}
  && [D^{--},  D^+_\a] =  D^-_\a\,, \label{24'}\\
  && \{ D^\pm_\a,  D^\pm_\b\} = 0\,. \label{24''}
\end{eqnarray}

It is easy to show that $\overline{W}$ is annihilated by the gauge covariant harmonic derivative:
\begin{eqnarray}
  {\nabla^{++}} \overline{W} &=& D^{++}(D^+)^2V^{--} + i[V^{++},\overline{W}]\nonumber\\
  &=& (D^+)^2(D^{--}V^{++} - i\left[ V^{++}, V^{--}\right])  + i[V^{++}, \overline{W}]\nonumber\\
  &=& -i[V^{++},\overline{W}] + i[V^{++},\overline{W}] = 0  \label{25}
\end{eqnarray}
where we have used the flat relations (\ref{24'}), (\ref{24''}) and $[D^{++}, D^+_\a]=0$, the defining equation (\ref{22'}) and the G-analyticity of $V^{++}$. Since $\overline{W}$ carries no U(1) charge, eq. (\ref{25}) implies that it is harmonic independent. Strictly speaking, this only applies to the Abelian curvature for which eq. (\ref{25}) becomes $D^{++}\overline{W} = 0$. The gauge covariant non-Abelian curvature $\overline{W}$ is only covariantly independent. However, the gauge invariant Lagrangian ${\cal L} = \mbox{Tr} (\overline{W} \overline{W})$ is simply harmonic independent, $D^{++}{\cal L} = 0$.

The covariant harmonic independence of $\overline{W}$ yields that it is annihilated by the other harmonic derivative as well,
\begin{equation}\label{25.1}
  {\nabla}^{--} \overline{W} = D^{--}\overline{W} + i[V^{--}, \overline{W}] = 0\,.
\end{equation}
This property is not derived by algebraic manipulations but rather follows from the fact that eq. (\ref{22'}) is a zero-curvature condition (see \cite{book} for  details). This means that both $V^{++}$ and $V^{--}$ can be written in the ``pure gauge" form $V^{\pm\pm} = -ie^{ib} D^{\pm\pm} e^{-ib}$. Here $e^{ib}$ is the analog of the $N=1$ $e^{iV}$, although it is a constrained superfield in the $N=2$ case. With its help one can gauge-rotate the curvature, $\overline{W}' = e^{ib} \overline{W} e^{-ib}$, so that the covariant equation (\ref{25}) becomes flat, $D^{++} \overline{W}' = 0 \ \Rightarrow\ D^{--} \overline{W}' = 0$.

Finally, the antichirality of $\overline{W}$ is partially manifest, since its expression (\ref{23}) satisfies $D^+_{\a} \overline{W} = u^+_iD^i_{\a} \overline{W} = 0$. However, this is only half of the antichirality condition, the other half takes a covariant form. Indeed, if in this scheme the flat spinor derivative $D^+_\a$ is gauge covariant and satisfies the commutation relation
\begin{equation}\label{25.2}
  [{\nabla}^{++}, D^+_\a] = 0 \quad \Leftrightarrow \quad  D^+_\a V^{++} = 0\,,
\end{equation}
the same is not true for $\nabla ^-_\a$ which is defined through the covariant version of eq. (\ref{24'}):
\begin{equation}\label{25.3}
  \nabla ^-_\a = [{\nabla}^{--}, D^+_\a]\,.
\end{equation}
So, hitting eq. (\ref{25.1}) with $D^+_\a$ and using (\ref{25.3}) we obtain
\begin{equation}\label{25.4}
  \nabla ^-_\a \overline{W}  = 0 \quad \Rightarrow \quad D^-_\a \overline{W} = 
  i[D^+_\a V^{--}, \overline{W}]\,.
\end{equation}
Once again, the  gauge invariant and harmonic independent Lagrangian ${\cal L} = \mbox{Tr} (\overline{W} \overline{W})$ is antichiral in the usual sense, $D^\pm_\a{\cal L} = 0 \ \Rightarrow D^i_\a {\cal L} = 0$.

\section{Deforming $N=2$ super-Yang-Mills}

If two superfields $A,B$ are antichiral, ${D}_{\a i}(A,B)=0$, or G-analytic, $D^+_\a(A,B)=\bar D^+_{\da}(A,B) = 0$, their star product (\ref{26}) is reduced to the usual one, $A\star B = A B$. This is however not true if one of the superfields does not have this property. In particular, if $A$ is G-analytic but $B$ is not,  their star commutator does not vanish:
\begin{equation}\label{27}
  [A,B]_\star = [A,B] -iP\left[D^{-\a}A, D^+_\a B\right] + \frac{P^2}{4} \left[(D^-)^2A, (D^+)^2 B\right]\,.
\end{equation}
To obtain (\ref{27}) we have projected ${D}_{\a i}$ in (\ref{26}) with harmonics. We have also taken into account the G-analyticity of the factor $A$, $D^+_\a A=0$. Note that expanding the exponential in (\ref{26}) we also obtain terms cubic or quartic in the spinor derivatives, e.g., $D^-(D^+)^2$ or $D^+(D^-)^2$, but they vanish when one of the factors is G-analytic. 

Let us apply this to the $N=2$ gauge theory. As before, we choose a G-analytic harmonic gauge connection $V^{++}(x_A,\q^+,\bq^+,u)$. The analog of the gauge transformation (\ref{19}) remains undeformed,
\begin{equation}\label{28}
  \d V^{++} = -D^{++}\Lambda + i[\Lambda, V^{++}]_\star = -D^{++}\Lambda + i[\Lambda, V^{++}]
\end{equation}
because both $\Lambda$ and $V^{++}$ are G-analytic. Consequently, the WZ gauge (\ref{20}) is still valid. However, the non-analytic gauge connection ${\cal V}^{--}$ is affected by the deformation:
\begin{equation}\label{29}
  \d {\cal V}^{--} = -D^{--}\Lambda + i[\Lambda, {\cal V}^{--}]_\star\,.
\end{equation}
Similarly, the defining differential equation (\ref{22'}) is deformed as well:
\begin{equation}\label{30'}
  D^{++}{\cal V}^{--} - D^{--}V^{++} + i\left[ V^{++}, {\cal V}^{--}\right]_\star = 0\,,
\end{equation}
or in detail (cf eq. (\ref{22})),
\begin{equation}\label{30}
  \nabla^{++}{\cal V}^{--} - D^{--}V^{++} +P \left[D^{-\a}V^{++}, D^+_\a {\cal V}^{--}\right]  + \frac{iP^2}{4} \left[(D^-)^2 V^{++}, (D^+)^2 {\cal V}^{--}\right] 
\end{equation}
where $\nabla^{++}$ is the undeformed gauge covariant harmonic derivative.

Further, trying to construct the deformed curvatures (\ref{23}) we realize that the expression for the chiral $W$ is not covariant anymore because $(\bar D^+)^2[\Lambda, {\cal V}^{--}]_\star \neq [\Lambda, (\bar D^+)^2{\cal V}^{--}]_\star$ (since $\{\bar D^+, D^-\}\neq 0$). This is of course a consequence of the fact that our star product (\ref{26}) breaks chirality. On the contrary, the antichiral $\overline{\cal W} = (D^+)^2 {\cal V}^{--}$ still is gauge covariant because $[\Lambda, \overline{\cal W}]_\star = [\Lambda, \overline{\cal W}]$ (both $\Lambda$ and $\overline{\cal W}$ are annihilated by $D^+$). We conclude that only the second of the two forms of the $N=2$ SYM action (\ref{21}), which were  equivalent in the undeformed case, can be used after the deformation (\ref{26}).

In order to find the deformed action, we still need to solve eq. (\ref{30}) for ${\cal V}^{--}$. We proceed as follows. Assume that we know the solution of eq. (\ref{22}) for the undeformed connection $V^{--}$. With its help we construct the undeformed antichiral curvature $\overline{W} = (D^+)^2 V^{--}$. Next, a straightforward but lengthy calculation, making use of eq. (\ref{22}), of the properties of $\overline{W}$ and of the various relations listed at the end of Section 3, shows that the following expression
\begin{equation}\label{32}
  {\cal V}^{--} = V^{--} + \frac{P}{4}\left[D^{+\a}V^{--}, \left\{D^+_\a V^{--},    \left(1 + P\overline{W} \right)^{-1} \right\}  \right] - \frac{iP^2}{4}\left[\overline{W} \left(1 + P\overline{W} \right)^{-1},  D^{+\a} D^-_\a V^{--}\right]
\end{equation}
is the (unique) solution of eq. (\ref{30}). Then, from eq. (\ref{32}) we easily obtain the deformed curvature
\begin{equation}\label{33}
  {\cal \overline{W}} = (D^+)^2 {\cal V}^{--} = \frac{{\overline{W}}}{1 + P{\overline{W}} } \ .
\end{equation}
Another direct calculation shows that it satisfies the deformed versions of eq. (\ref{25.1})
\begin{equation}\label{25.1'}
  {\nabla}^{--}_\star {\cal \overline{W}} = D^{--}{\cal \overline{W}} + i[{\cal V}^{--}, {\cal \overline{W}}]_\star = 0\,,
\end{equation}
and of its corollary (\ref{25.4}).

Finally, the deformed action reads
\begin{equation}\label{34}
  S^{N=2}_{\rm deformed\ SYM} =\frac{1}{64}\int d^4x_R\, d^4\bq\; {\rm Tr}\; {\cal \overline{W}}^2 =\frac{1}{64}\int d^4x_R\, d^4\bq\;  {\rm Tr} \left( \frac{{\overline{W}}}{1 + P{\overline{W}} }  \right )^2\,.
\end{equation}
Note that the product used in (\ref{34}) is the ordinary one, ${\cal \overline{W}}\star {\cal \overline{W}} = {\cal \overline{W}}{\cal \overline{W}}$,  since ${\cal \overline{W}}$ and ${\overline{W}}$ are annihilated by $D^+_\a$. Also, the action (\ref{34}) is manifestly invariant under the full $N=2$ supersymmetry. So, the only effect of our deformation is the particular non-minimal gauge invariant coupling in (\ref{34}). 

This action resembles the so-called ``special geometry" actions \cite{dWLvP,CFD,St} 
\begin{equation}\label{35}
  S = \int d^4x_L\, d^4\q\;  f(W) + \int d^4x_R\, d^4\bq\;  \bar f(\overline{W})
\end{equation}
where $f(W)$ is a holomorphic gauge invariant function of the chiral curvature $W$. The main difference is that in (\ref{34}) we only see the antiholomorphic term (recall that our deformation does not allow us to construct $W$). Clearly, our action is not real, but this violation of reality just reflects the choice of the star product (\ref{26}) and is also observed in the $N=1$ case \cite{S}.

Let us examine the nature of the interaction terms contained in the deformed action. For simplicity we restrict ourselves to the Abelian case only, the full non-Abelian component action can be found in \cite{dWLvP}. We first expand $\overline{W}$ (which now is the standard Abelian $N=2$ curvature) in terms of the component fields (\ref{20}): 
\begin{equation}\label{36}
  \overline{W}(x_R,\bq) = \bar\phi(x_R) + \bq_{\da i} \bar\lambda^{\da i} + \bq_{i}\tilde\s^{\m\n}\bq^i F^-_{\m\n} + \bq_i\bq_j D^{ij} + (\bq^3)^i_{\da} (\tilde\s^\m)^{\da\a} i\pa_\m \lambda_{\a i} - (\bq)^4 \square\phi\,,
\end{equation}
where $F^-_{\m\n}$ is the anti-selfdual part of the Abelian field strength.
Introducing the antiholomorphic function $\bar f(\bar\f) = \bar\f^2 (1+P\bar\f)^{-2} = \bar\f^2 - 2P\bar\f^3 + 3 P^2 \bar\f^4 + O(P^3)$, we find the component Lagrangian
\begin{eqnarray}
  {\cal L}&=& -\frac{1}{2}\bar f'(\bar\f)\square\phi -\frac{1}{4}\bar f''(\bar\f) (F^-_{\m\n})^2 -\frac{1}{2}\bar f''(\bar\f)\bar\lambda_i\s^\m i\pa_\m \lambda^i + \frac{1}{8}\bar f''(\bar\f) D^{ij}D_{ij}\nonumber\\
  &&  + \frac{1}{2}\bar f'''(\bar\f)\bar\lambda_{\da i} \bar\lambda_{\db j} \left[   \e^{ij} (\tilde\s^{\m\n})^{\da\db} F^-_{\m\n} + \e^{\da\db} D^{ij}\right] + \bar f^{ iv}(\bar\f) (\bar\lambda)^4\,.  \label{37}
\end{eqnarray}
In the $N=1$ Abelian case \cite{S} the only deformation term has the form $\bar\lambda_{\da}\bar\lambda_{\db}\e^{\da\db} C^{\m\n}F^+_{\m\n}$. Our term $\bar\lambda_i\tilde\s^{\m\n}\bar\lambda_j \e^{ij} F^-_{\m\n}$ can only exist in $N=2$, since it is made out of two fermion fields, $\bar\lambda_1\bar\lambda_2$. In addition, here we have some new, purely bosonic terms. Let us examine the corresponding field equations in which we set the fermions to zero for simplicity:
\begin{eqnarray}
  \d\f:&&\square\left[\bar f'(\bar\f) \right]=0 \nonumber\\
  \d\bar\f:&&\square\phi = -\frac{\bar f'''(\bar\f)}{2\bar f''(\bar\f)} (F^-_{\m\n})^2  \label{38}  \\
  \d A^\n:&& \pa^\m \left[\bar f''(\bar\f)F^-_{\m\n}\right] = 0      \nonumber\\
  \d D_{ij}:&& D^{ij} = 0\,.  \nonumber
\end{eqnarray}
It is clear that after the field redefinition $\bar\varphi = \bar f'(\bar\f)$ the scalar $\bar\varphi$ becomes a free field (this is true even in the presence of the fermions). However, the scalar $\f$ has a source proportional to the deformation parameter $P$.\footnote{It is important to reiterate that in the Euclidean superspace formulation that we are using, the fields $\phi, \bar\phi$ and  $\lambda, \bar\lambda$ are not related to each other by complex conjugation, but should be treated as independent fields.} Similarly, the right-handed spinor remains free (after the field redefinition $\bar\tau_{\da i} = \bar f''(\bar\f)\bar\lambda_{\da i}$), but the left-handed $\lambda$ interacts with the other fields. This suggest to use the term ``heterotic special geometry" to characterize the deformed action (\ref{34}).

\section{Conclusions}

In this paper we have exploited the unique possibility of deforming $N=2$ superspace with a singlet scalar parameter $P$ and have constructed the corresponding deformed non-Abelian gauge theory. Our Moyal-Weyl star product is manifestly supersymmetric. Although the deformation does not affect the G-analytic gauge superfield and its WZ gauge, it manifests itself in the form of non-polynomial corrections to the antichiral curvature. In this way we have demonstrated that it is possible to have a deformed action which preserves the full $N=2$ supersymmetry.

We remark that the coupling of the $N=2$ SYM gauge superfield $V^{++}$ to the hypermultiplet matter superfield $q^+$  (\ref{17}), (\ref{18}) remains undeformed since both superfields are G-analytic, so $[V^{++} , q^+]_\star = [V^{++} , q^+]$. It is well known that such a gauge-matter system is equivalent to the $N=4$ SYM theory. In other words, the action $S^{N=2}_{{\rm matter}} + S^{N=2}_{{\rm SYM}}$ has two extra non-linear supersymmetries which transform $V^{++}$ and $q^+$ into each other \cite{book}. Our deformation affects only the $N=2$ gauge sector, so it is likely to break the $N=4$ supersymmetry down to $N=2$.

Finally, a word about the possibility to employ the Lorentz tensor and  SU(2) triplet deformation parameter $P^{\m\n (ij)}$ from eq. (\ref{1}). If we still insist on using the $D$ star product (and thus on preserving the full supersymmetry), we are going to break G-analyticity. Indeed, if both $A$ and $B$ are G-analytic, in $A\star B$ we can have terms like $P^{\a\b++}D^-_\a A D^-_\b B$ which are not G-analytic. Since G-analyticity is crucial for constructing the $N=2$ SYM action, the way out could be to give up half of the $N=2$ supersymmetry by using $\eth = iQ$ in the star product (following \cite{S}). This alternative deformation is under investigation. 

\bigskip {\bf Note added.} After the first version of this paper had been submitted to the e-archive we became aware of the recent paper \cite{ILZ} where the deformations of harmonic superspace, in particular the $Q$ deformation with a singlet scalar parameter, are discussed.

\section*{Acknowledgements}
The work of S.F. and E.S. has been supported in part by the INTAS grant No 00-00254 and that of S.F. by the D.O.E. grant DE-FG03-91ER40662, Task C and by the European Community's Human Potential Program under contract HPRN-CT-2000-00131 ``Quantum Space-Time". E.S. acknowledges discussions with Evgeny Ivanov.

\end{document}